\documentclass[%
 reprint,
superscriptaddress, amsmath,amssymb,
]{revtex4-2}

\usepackage{graphicx}
\graphicspath{{Figures/}{}}

\usepackage{dcolumn}
\usepackage{physics}
\usepackage{siunitx}
\AtBeginDocument{\RenewCommandCopy\qty\SI}
\usepackage{bm}

\usepackage[pdfstartview={FitH},pdfpagemode={UseNone}, breaklinks=true,colorlinks=true,bookmarks=false,allcolors=blue,bookmarksopen=true, pdfnewwindow=true]{hyperref}
\usepackage[all]{hypcap}

\usepackage[english]{babel}

\begin{document}

\title{Full conversion of unpolarized to fixed-polarization light \\with topology optimized metasurfaces}

\author{Neuton Li}
  \email{Neuton.Li@anu.edu.au}
\author{Jihua Zhang}
\affiliation{ARC Centre of Excellence for Transformative Meta-Optical Systems (TMOS), Department of Electronic Materials Engineering,  Research School of Physics, Australian National University, Canberra, ACT 2601, Australia}

\author{Shaun Lung}
\affiliation{ARC Centre of Excellence for Transformative Meta-Optical Systems (TMOS), Department of Electronic Materials Engineering,  Research School of Physics, Australian National University, Canberra, ACT 2601, Australia}
\affiliation{Institute of Applied Physics, Friedrich-Schiller University, 07743 Jena, Germany}

\author{Dragomir N. Neshev}
\author{Andrey A. Sukhorukov}%
\affiliation{ARC Centre of Excellence for Transformative Meta-Optical Systems (TMOS), Department of Electronic Materials Engineering,  Research School of Physics, Australian National University, Canberra, ACT 2601, Australia}

\date{21st August 2023}

\begin{abstract}
Conventional polarizers and polarization beam splitters have a fundamental limit of 50\% efficiency when converting unpolarized light into one specific polarization.  
Here, we overcome this restriction and achieve near-complete conversion of unpolarized light to 
a single
pure polarization 
state
at several outputs of 
topology-optimized metasurfaces.
Our fabricated metasurface achieves 
an extinction ratio approaching $10^2$, when characterized with laboratory measurements. We further demonstrate that arbitrary power splitting can be achieved between three or more polarized outputs, offering flexibility in target illumination. Our results provide a path toward greatly improving the efficiency of common unpolarized light sources  in a variety of applications requiring pure polarizations. 
\end{abstract}

\maketitle


Polarization is a fundamental property of light that can carry and probe information with a wide range of applications including imaging~\cite{Ramella-Roman2020AApplications, Rowe1995Polarization-differenceMedia, Nan2009LinearApplications}, sensing~\cite{Yan2020GeneralSensing,Yan:2020:PolarizationSensing}, and communications~\cite{Damask:2004:PolarizationTelecommunications, Guo2017AdvancesApplications}, which often benefit from having a predefined pure polarization state as the input or output. 
Whereas lasers with a polarized output are available, the cheaper and ubiquitous sources such as light-emitting diodes usually emit unpolarized or partially-polarized light, and so efficient extraction of fully polarized light from such sources remains a challenging problem~\cite{Matioli:2012-e22:LSA, Wang:2020-2001890:ADM, Huang:2023-111107:APL}. The traditional approach is to
filter out the undesired orthogonal state by a linear polarizer, in combination with waveplates if the target is an elliptical or circular polarization state~\cite{Chekhova:2021:PolarizationLight}.
This process is energy inefficient since
there is a \SI{50}{\percent} limit for conversion of unpolarized to fully polarized light with conventional polarizers~\cite{Collett:2005:GuidePolarization}. 
An energy-efficient approach could be based on the polarization beam splitter (PBS), which directs orthogonal linear polarizations into separate paths. In this way, the total transmitted power is conserved, however, the resulting beams do not have the same polarization. In order to obtain the desired  polarization, additional waveplates are required adding bulk and complexity to the system, which would be incompatible with compact end-user devices requiring integrated optics solutions.

\begin{figure}[b]
    \centering
    \includegraphics[width = 0.95\linewidth]{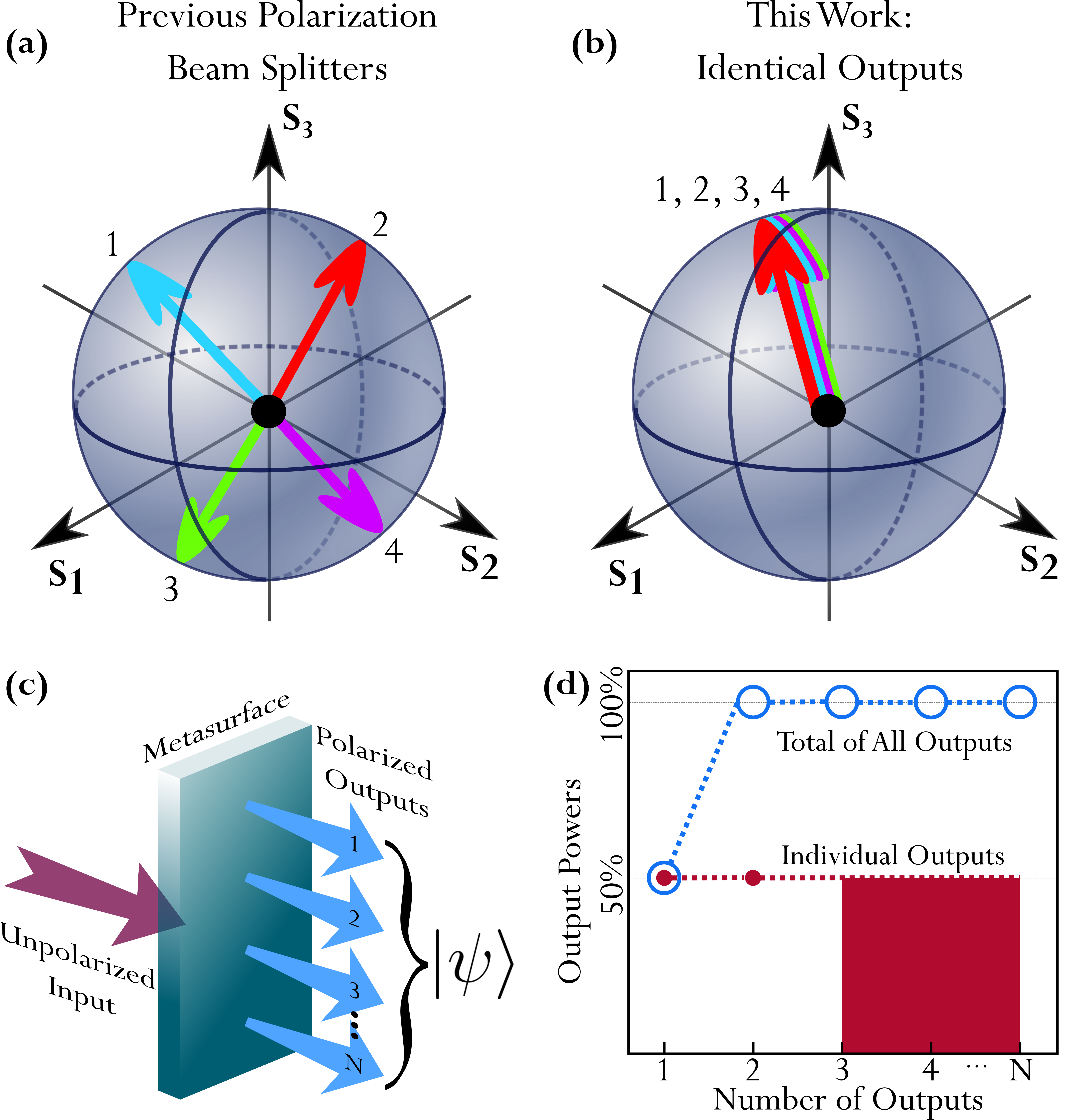}
    \caption{(a,b)~Output polarization states shown on a 
    Poincar\'e sphere for (a)~previously realized metasurfaces that split polarizations in pairwise orthogonal states between different outputs  (1 and 4, 2 and 3) and (b)~the proposed metasurface which achieves identical output polarization states.
    (c)~A schematic depicting a metasurface which converts an unpolarized input into $N$ outputs with identical polarization states $\ket{\psi}$. (d)~The permissible transmitted power of an ideal polarizer for a different number of outputs. The red markers show the allowed transmission of individual output channels, and the blue markers show the maximum total transmission of all output channels. }
    \label{fig:1}
\end{figure}

In the last decade, there were great advances in shaping polarization states of light with optical metasurfaces, composed of a planar array of nanostructures with a subwavelength thickness \cite{Kildishev2013PlanarMetasurfaces, Quevedo-Teruel2019RoadmapMetasurfaces}. In polarization manipulation, metasurfaces exploit the shape birefringence supported by the anisotropic nanostructures, which can be tailored to manipulate a single polarization or a pair of orthogonal polarizations independently \cite{Lung2020Complex-BirefringentTransformations, BalthasarMueller2017MetasurfacePolarization, Cerjan2017AchievingMetamaterials}, with arbitrary linear, elliptical, or circular shape of polarization ellipse. Because of this, a single metasurface can enable arbitrary polarization conversion and dichroism to realize the functions of 
waveplates and 
polarizers, respectively. 
For example, arbitrary polarization manipulation conversion was demonstrated with single-layer dielectric metasurfaces~\cite{Wang2021ArbitraryPolarizers, Ding2020VersatileMetasurfaces, Hu:2020-3755:NANP}. A metasurface hologram combining different polarizing meta-atoms into an array can be used as a polarimeter by projecting different states into spatially distinct regions \cite{Zhang:2019-1190:OPT, Arbabi2018Full-StokesMetasurfaces, Gao2021EfficientControl, Rubin2021JonesMetasurfaces}.
In addition, recent work has demonstrated the ability to control the degree of polarization with a metasurface, by filtering unpolarized incident light to arbitrary partially or fully polarized output~\cite{Wang2023Metasurface-BasedPolarizer}. Metasurfaces have also replicated and extended the functionality of a conventional PBS~ \cite{Zhang2018DesignMetasurface, Wang2018Rochon-Prism-LikeMetasurfaces, Khorasaninejad2015EfficientMetasurface, Slovick2017MetasurfaceSplitter, Rubin2021JonesMetasurfaces}, by splitting incoming light into multiple pairs of orthogonally polarized beams as illustrated on a Poincar\'e sphere in Fig.~\ref{fig:1}(a). 
With these devices, the total energy transmitted through the metasurface is conserved in split beams of different polarizations. However, such metasurfaces still rely on the same principle as conventional bulky PBS, that is separating orthogonal polarization components of the light into different directions. Therefore, they are still subject to the fundamental limit of 50\% in the conversion of unpolarized light to one specific fully polarized output state. If this limit were to be surpassed, greater efficiencies will be achieved in optical devices, especially in shaping light from unpolarized sunlight, thermal lamp, and light emitting diodes~\cite{Kim2022Metasurface-drivenImaging}.

\begin{figure*}[t]
    \centering
    \includegraphics[width =0.95\linewidth]{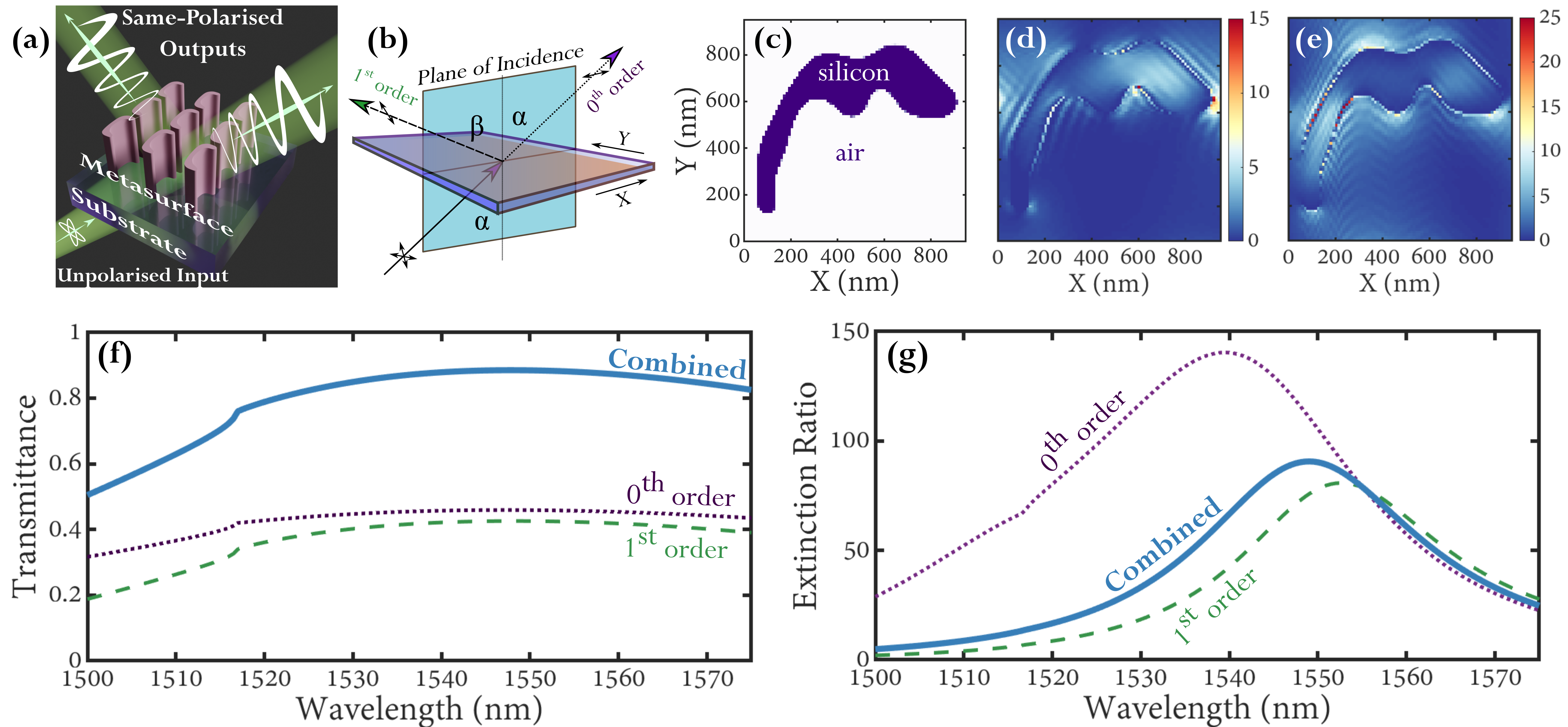}
    \caption{(a) Illustration of the dual-output polarizing metasurface (b)~Operating scheme of the metasurface in transmission, with an incident angle of $\alpha=\SI{45}{\degree}$ and first order diffraction angle of $\beta=\SI{68}{\degree}$. (c) Final metasurface design of single unit cell, with a period \SI{950}{\nano\metre} along both directions. The shaded region represents silicon, and void represents air. (d) Electric field intensity enhancement for incident $\ket{H}$ polarisation, and (e) $\ket{V}$ polarisation. These have the same unit cell dimensions as (c). (f) Predicted transmitted power, and (g) extinction ratio of the diagonal linear polarisation for each diffraction order. }
    \label{fig:2}
\end{figure*}

In this work, we reveal that ultimate efficiency and flexibility in converting unpolarized to a single polarization state can be achieved through specially designed elements with multiple output channels, which overcome the efficiency limit of a single-output polarizer (Fig.~\ref{fig:1}(b)). 
We implement this principle by inversely designing metasurfaces with two, three, and four outputs, which can convert unpolarized light into a single predefined output polarization state with combined efficiency far exceeding the 50\% threshold. In experiments, we demonstrate the dual-output metasurface polarizer, with the measured efficiency reaching 70\%.
These fundamental advances in polarization optics can improve the energy efficiency of many optical technologies employing unpolarized or partially polarized light sources.

We first formulate the general properties of any passive linear optical device with $N$ total outputs, where an unpolarized light source is coupled to the single input port (Fig.~\ref{fig:1}(c)). 
For maximum efficiency of unpolarized to polarized light conversion, we aim to fully transmit all the input power across the outputs, with each output having a predefined target pure polarization state $|\psi_{1,2,...,N}\rangle$. 
Mathematically, such a transformation to an arbitrary output state $\ket{\psi_n}$ of the $n^{th}$ output can be defined by a Jones matrix
\begin{equation} \label{eq:Jn}
    \mathbf{J}_n = A_n |\psi_n\rangle \langle \phi_n| ,
\end{equation}
where 
$|\phi_n\rangle$, the input state, is optimized to maximize the transmission efficiency, and
$A_n$ are the real-valued transmission amplitudes that are bounded for passive devices as $0 \le A_n \le 1$. An input unpolarized light source can be represented as a mixed state with the density matrix
\begin{equation} \label{eq:rhoUnpol}
    \rho_{\rm in}=\frac{1}{2}(|H\rangle \langle H| + |V\rangle \langle V|).
\end{equation}
The corresponding power transmission to each output is then found as $P_{{\rm unpol}, n} = 0.5|A_n|^2$. This means that the maximum conversion efficiency for each output is 50\% (red dotted line in Fig.~\ref{fig:1}(d)), as for a conventional polarizer with a single output. The total power efficiency is $P_{{\rm unpol, total}} = 0.5\sum_{n=1}^N |A_n|^2$, which should be no more than one for passive devices. Then, the general research question becomes: can we find a set of $|\phi_{1,2,...,N}\rangle$ such that $P_{{\rm unpol, total}}=1$ for any given set of output states $|\psi_{1,2,...,N}\rangle$ and arbitrary power splitting portions to each output $P_{{\rm unpol}, n}$.

We find that the full power efficiency can be achieved for $N \geq 2$, as shown by the blue markers in Fig.~\ref{fig:1}(d) and see Supplementary~S1 \cite{SM} for furtherdetails. For $N=2$, 
the power transmission to each of the two outputs is exactly 50\%, meaning that we do not have flexibility in the power splitting when realizing 100\% total efficiency. Nevertheless, in this scenario one can still have an arbitrary choice of the pure output polarization states, including having the same output polarization states. In comparison, previous bulky-optics and metasurface based PBS have never explored the full potential of a dual-output polarizer. 

For three or more outputs, 100\% total efficiency can be achieved with arbitrary output states and any power splitting portions, only subject to a condition that each output has a maximum of 50\% power, as marked by the red shading in Fig.~\ref{fig:1}(d). After defining the output states and power splitting portions, analytical forms of the right singular states $|\phi_{1,2,3}\rangle$ can be obtained for three outputs which are derived in Supplementary~S1.3 \cite{SM}. For $N\ge4$ output ports, there are non-unique solutions since the number of allowed free parameters over-satisfy the necessary conditions, and we present a particular analytical solution in Supplementary~S1.4 \cite{SM}. We emphasise that this is a general result; a passive optical device with $N\geq2$ ports may have 100\% total efficiency in achieving arbitrary output polarization states. In addition, a device with $N\geq3$ ports may also have arbitrary power splitting portions at the outputs, provided that no single port contributes more than 50\% to the total efficiency. To replicate the functionality of this device with a conventional optical system would require a series of multiple PBSs, waveplates, and other bulky optical elements.

\begin{figure*}[t]
    \centering
    \includegraphics[width = 0.99\linewidth]{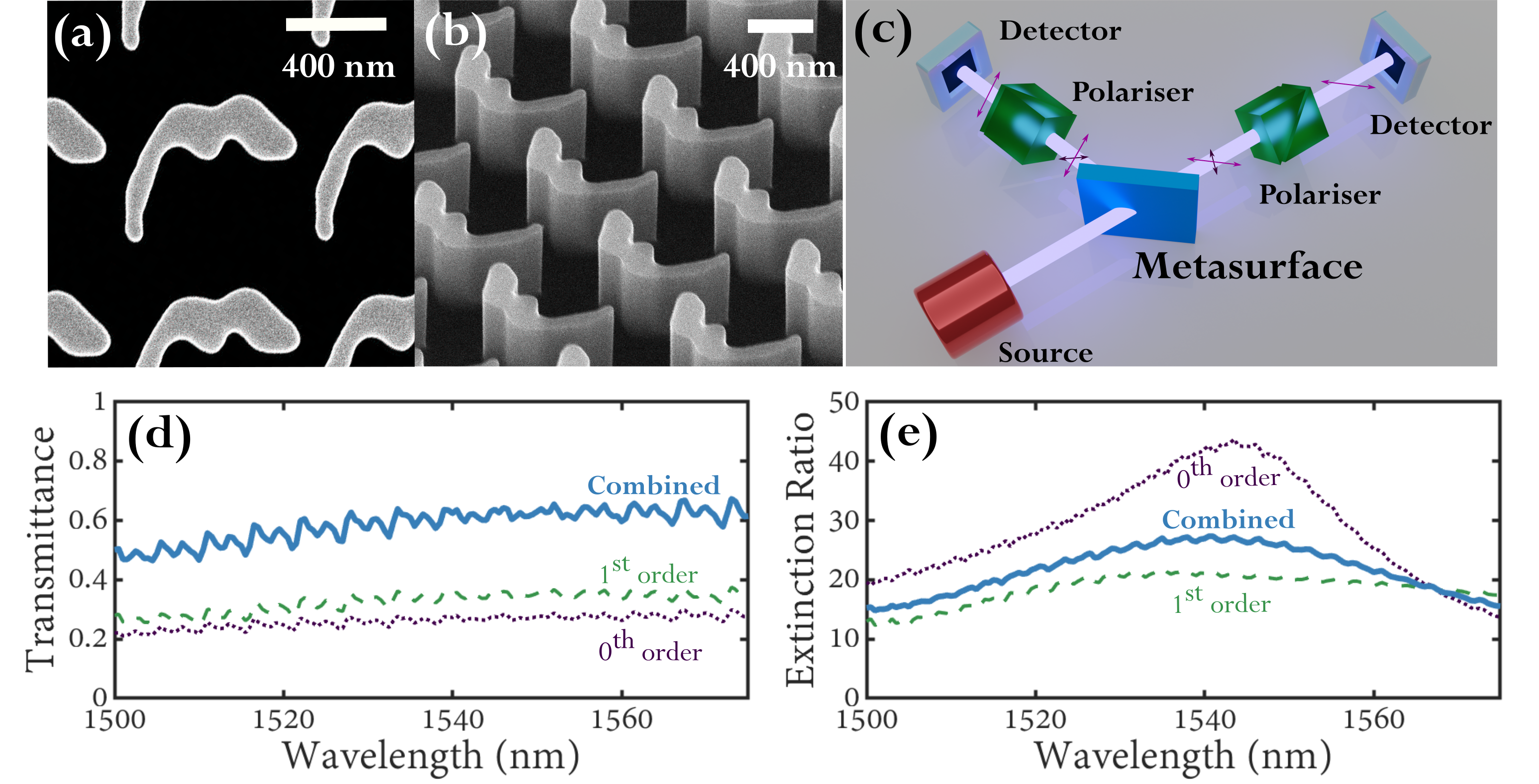}
    \label{fig:SEM_experiment}
    \caption{(a) Top-down and (b) tilted SEM images of fabricated metasurface. (c) Simplified experimental setup. The source transmits through the metasurface, which diffracts the beam into two orders. Each arm analyses the diagonal polarization before reaching the detector. (d) Measured transmitted power, and (e) extinction ratio of the diagonal polarization for each diffraction order for an incident angle $\alpha=\SI{45}{\degree}$. }
\end{figure*}

Next, we illustrate a particular case of polarization conversion by designing meta-gratings that split an incoming unpolarized beam into multiple diffraction orders, all having identical output pure polarizations $|\psi_{1,2,...,N}\rangle = |\psi\rangle$. This operational functionality can be beneficial for structured illumination applications~\cite{Ni:2020-6719:NANL}. In these scenarios, converting the generated light into specific output polarization can be used to reduce unwanted reflections from transparent surfaces. 

Dielectric metasurfaces are commonly designed in the framework of weakly-interacting uncoupled resonators \cite{Rubin2021JonesMetasurfaces, Kamali2018AControl, Chen2020FlatMetasurfaces, Yu2014FlatMetasurfaces, Qiu2021FundamentalsMetasurfaces}, where the output fields are determined by the superposition of position-dependent transfer matrices. These compositions are very successful in many applications, however, we find that this type of metasurface construction does not enable the desired functionality of our research task. Specifically, according to Eq.~(6) in Ref.~\cite{Rubin:2021-eabg7488:SCA} $J_n = A_n\ket{\phi_n^\ast}\bra{\phi_n}$, conventional non-chiral pillar-based metasurfaces are restricted to the target Jones matrices with $|\psi_{n}\rangle = |\phi_n^\ast\rangle$. It means that for identical output pure polarizations $|\psi_{n}\rangle = |\psi\rangle$, one necessarily has $|\phi_n\rangle = |\psi^\ast\rangle$, such that the device would filter out one input polarization component with the same output efficiency limit of 50\% as a conventional polarizer.

We overcome this apparent roadblock by designing dielectric metasurfaces with spatially non-local response, where the polarisation transformations depend on the diffraction order. In principle, as each diffraction order can permit an independent transfer matrix, full conversion of unpolarized light to a specific polarisation state can be possible with an ideal design. For this purpose, we perform inverse design with free-form topology optimization \cite{Fan:2020-196:MRSB, Shi:2020-eaba3367:SCA, Jensen2011TopologyNano-photonics, Li2022EmpoweringApplications} by adopting the MetaNet codebase~\cite{Jiang:2020-13670:OE} in combination with RETICOLO rigorous coupled wave analysis (RCWA) electromagnetic solver~\cite{Hugonin:2101.00901:ARXIV}. During the optimization, the figure of merit (FOM) is formulated as the difference between the transmitted target and undesired orthogonal polarisation intensities, multiplied over all the output diffraction orders:
\begin{equation}
    FOM = \prod_n \left[ \abs{\bra{\psi}\textbf{J}_n}^2 - \abs{\bra{\psi_{\perp}}\textbf{J}_n}^2 \right] ,
    \label{Eq:FOM}
\end{equation}
where $\ket{\psi}$ is the desired output polarization state for all diffraction orders, and $\ket{\psi_\perp}$ is an undesired orthogonal state ($\bra{\psi}\ket{\psi_\perp}=0$). Then, by maximizing this function, the optimized device approaches the intended operation. We calculate the function gradient through adjoint simulations, allowing for fast convergence of iterations.

In the first design, we target the splitting of an incoming unpolarized beam into two outgoing beams ($N=2$) with the same diagonal linear polarization $|D\rangle$. 
This can be achieved with an angled incidence of the input beam such that only two diffraction orders exist in the transmission direction, as sketched in Fig.~\ref{fig:2}(a). 
We design the metasurface for equal transmitted diffraction efficiency into the $m=0$ and $m=1$ orders, as sketched in Fig.~\ref{fig:2}(b). 

The incident angle is $\alpha = 45^\circ$ for the unpolarized input light. 

We run the topology optimization  starting with a random distribution of refractive indices in the unit cell. The design is based on a silicon layer ($n=3.48$ at $\lambda=\SI{1550}{\nano\metre}$) with a thickness of \SI{1000}{\nano\metre} on \SI{460}{\micro\metre} sapphire substrate ($n=1.75$ at $\lambda=\SI{1550}{\nano\metre}$), corresponding to our physical sample. The algorithm not only maximizes the figure of merit, but includes binarization of the perturbation region to either silicon or air with a constraint on minimum feature size \cite{Wang2019RobustMetasurfaces}. The optimized complex shape of the nanoresonator is shown in Fig.~\ref{fig:2}(c). Due to its asymmetric shape, the fields induced in the resonator are highly non-trivial. For example, when the incident light is polarized in the \textit{X}-direction ($\ket{H}$ polarization) (Fig.~\ref{fig:2}(d)), we see that most of the field is concentrated on the right tip of the nanoresonator. However, when incident light is polarized in the \textit{Y}-direction ($\ket{V}$ polarization) (Fig.~\ref{fig:2}(e)), the light is generally more concentrated along the left arm of the nanoresonator. Therefore, the action of the metasurface on unpolarized light is defined through a complex superposition of the fields induced in the nanoresonator, which is arrived at by the optimization algorithm. We also performed multipolar decomposition of the fields to gain further insight into the underlying mechanism of the nanoresonator as shown in Supplementary~S2 \cite{SM} (see also reference \cite{Alaee2018AnApproximation} therein). We observe that the predominant modes in the nanoresonator are dipole modes. The electric and magnetic dipole modes have approximately equal scattering cross-sections, and are roughly an order of magnitude larger than their respective quadrupole counterparts. There are also no strongly resonant modes present in the nanoresonator, which enables the broadband operation of the optimized metasurface.

The modeling predicts the highly efficient conversion of unpolarized light to the target diagonal state $|D\rangle$ at both outputs (Fig.~\ref{fig:2}(f)). With Jones matrix analysis presented in Supplementary~S3 \cite{SM}, it is observed that both $\ket{H}$ and $\ket{V}$ polarizations are converted with near equal efficiencies to $\ket{D}$ polarization for each of the outputs. 
In simulations, each diffraction order reaches approximately 40\% power transmission at the target polarization state, for a combined total efficiency beyond 80\% over an extended wavelength range in the telecommunication band (\numrange{1540}{1560} \si{\nano\metre}). Importantly, this performance fundamentally exceeds the 50\% limit of a conventional polarizer. At the same time, the undesired orthogonally polarized state $|A\rangle$ has much lower power, with 
the combined extinction ratio approaching 100 at \SI{1550}{\nano\metre} (Fig.~\ref{fig:2}(g)). It is observed that the metasurface maintains effective performance greater than the threshold 50\% efficiency over the entire incident angle ($\alpha$) range from \SI{40}{\degree} to \SI{80}{\degree}. Peak conversion efficiency of $\sim$80\% is reached at around \SI{50}{\degree}  incidence which is shown in Supplementary~S4 \cite{SM}. The total efficiency can likely be increased beyond 90\% by introducing multi-layer metasurfaces or volumetric metamaterials, which have already demonstrated enhanced performance in devices for other applications \cite{Roques-Carmes2022TowardMetaoptics, Mansouree2020MultifunctionalOptimization, Shi:2022-167403:PRL, Roberts20233D-patternedMetaoptics}.  

We successfully fabricate the optimized design using e-beam lithography and standard silicon etching, as shown in Figs.~\ref{fig:SEM_experiment}(a,b). The characterization of the metasurface was then performed in free space by mounting it on a rotation stage in the horizontal plane to measure the $m=0$ and $m=1$ diffraction orders, as illustrated in Fig.~\ref{fig:SEM_experiment}(c).

The experiments were performed with the horizontal and vertical input polarization states in two separate measurements, mimicking the unpolarized incident light according to Eq.~(\ref{eq:rhoUnpol}).

These states were prepared from a continuous-wave tuneable laser, operating in the near-infrared within the telecommunications band wavelengths of \SIrange{1500}{1575}{\nano\metre} and collected in free-space via lenses. As given by Eq.~(\ref{eq:rhoUnpol}), unpolarized light can be formed by a combination of orthogonal states. These orthogonal input polarization states for the experiment are vertical $\ket{V}$ and horizontal $\ket{H}$, which are prepared and projected through the metasurface. For each \textit{input} polarization state, the corresponding \textit{output} polarization states were analyzed at each diffraction order by measuring the powers in the diagonal $\ket{D}$ and anti-diagonal $\ket{A}$ polarization components. By combining results, we may effectively determine the response of the metasurface to unpolarized light. For example, $P_{D,V_0}$ is defined as measuring the power output projected to $\ket{D}$ with input $\ket{V}$ for the zeroth order. We then infer the  power for $\ket{D}$ polarization in the zeroth order for an unpolarized input by averaging the measurements as $P_{D_0} = \frac{1}{2}(P_{D,V_0} + P_{D,H_0})$, and this is similarly done for the first order $P_{D_1}$. Calibration measurements, described in Supplementary~S5 \cite{SM}, against air are taken to determine the variation in input power across all relevant wavelengths. These values are then used to calculate the \textit{absolute} metasurface transmission efficiency.

We experimentally demonstrate the metasurface's ability to convert unpolarized light to diagonal polarised light with an efficiency close to $\sim$70\% [Fig.~\ref{fig:SEM_experiment}(d)], exceeding the 50\% limit of previous approaches. This performance is maintained across a broad wavelength range, from \SI{1540}{\nano\metre} to \SI{1570}{\nano\metre}. The extinction ratio of desired to undesired output polarization states exceeds 20 at the target wavelength of \SI{1550}{\nano\metre}, see Fig.~\ref{fig:SEM_experiment}(e). For another demonstration, we optimize for and experimentally characterize a two output circular polarizer, see Supplementary~S6 \cite{SM}. The metasurface is able to realize conversion of unpolarized light into circularly polarized light with approximately 60\% efficiency, thereby exceeding the 50\% limit. The performance of the experimental sample is lower than the simulations, which may be caused by significant reflection from the backside of the substrate which was not accounted for in the device simulations. We anticipate that experimental performance can be further improved by introducing anti-reflection coatings.

\begin{figure}[t]
    \centering
    \includegraphics[width =0.95\linewidth]{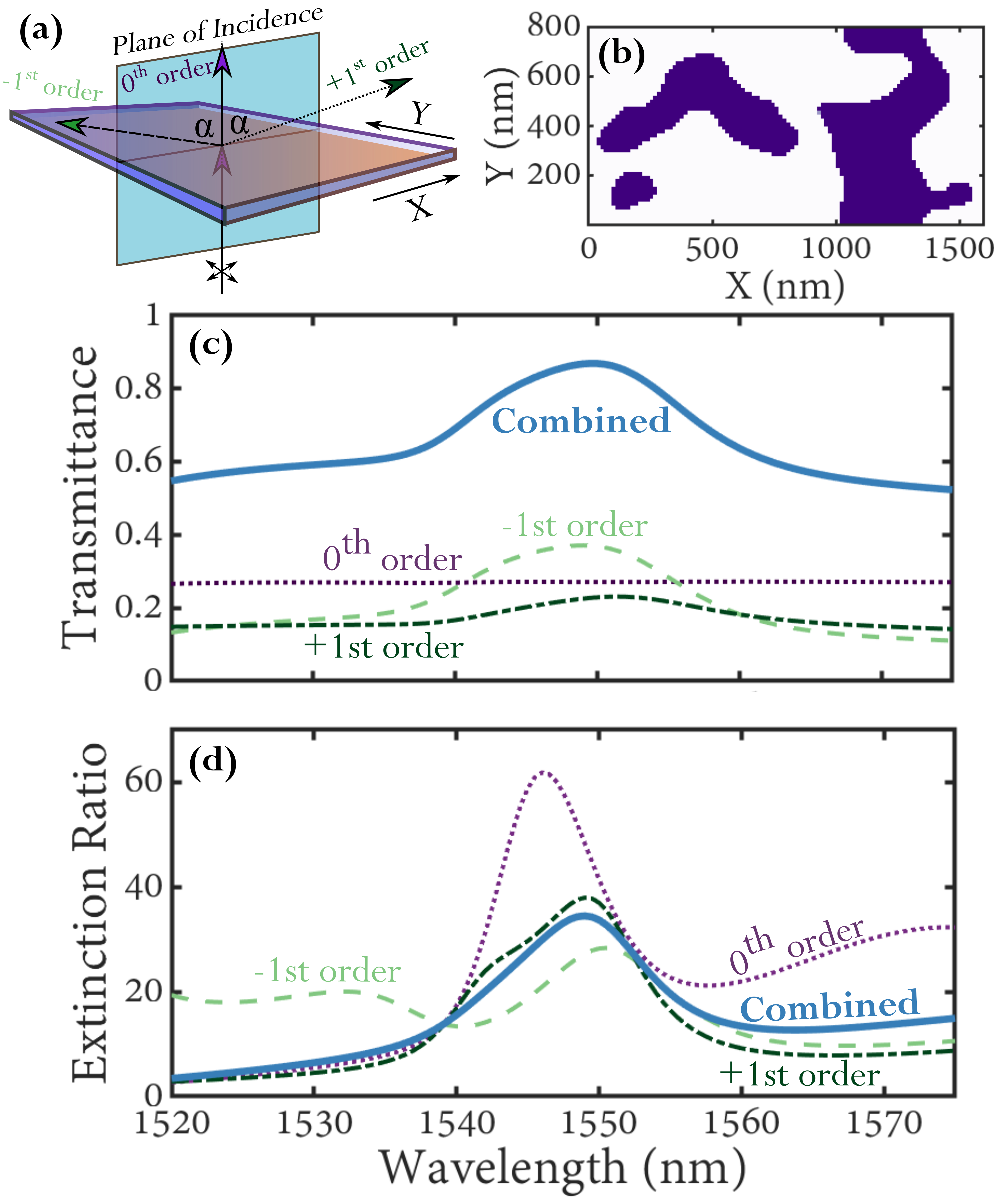}
    \caption{(a) Operating principle of a three-output metasurface polariser. The incident beam is normal, and $\alpha = \SI{75}{\degree}$. (b) The binarized metasurface pattern, with shaded regions representing silicon, and void representing to air. (c) Transmittance of desired vertical polarisation versus wavelength. (d) Extinction ratio of combined vertical to horizontal polarization. }
    \label{fig:3Out}
\end{figure}

Our method can then be extended to metasurface designs that generate three or more outputs. As discussed above and visualized in Fig.~\ref{fig:1}(d), there is freedom in the distribution of output powers for more than two diffraction orders. We present a metasurface with non-equal power splitting portions for each of the three outputs
(Fig.~\ref{fig:3Out}(a)), and the optimized design is shown in Fig.~\ref{fig:3Out}(b). With this 3-output design, high combined transmittance of over 80\% to the target vertical linear polarization $|V\rangle$ is predicted at \SI{1550}{\nano\metre}, while the extinction ratio over the $|H\rangle$ output approaches 40 at the same wavelength  (Fig.~\ref{fig:3Out}(c,d)). 
Moreover in Supplementary~S7 \cite{SM}, we find a nontrivial metasurface design where incoming unpolarized light is split into four outputs with the same polarization, while achieving similar performance.

In conclusion, we anticipate that the metasurfaces facilitating highly-efficient transformation and shaping of unpolarized light into fully polarized outputs can find various applications, including glare reduction in structured light illumination and polarized imaging with basic unpolarized sources such as LED and multi-mode lasers.
Our results also demonstrate that a nontrivial combination of local and nonlocal resonances in topologically optimized metasurfaces can overcome the limitations associated with arrays of weakly coupled resonators, and thereby open a path to broader polarization manipulation functionalities.

We acknowledge the support by the Australian Research Council (LP180100904, CE200100010).
This work was performed in part at the Melbourne Centre for Nanofabrication (MCN) in the Victorian Node of the Australian National Fabrication Facility (ANFF).

\newpage
\bibliography{references,references_mendeley}

\end{document}